\documentclass[twocolumn,showpacs,prb]{revtex4}
\usepackage{graphicx,dcolumn,bm,amsmath,amssymb,latexsym,float}

\begin{document}
  
\newcommand{\be}{\begin{equation}}
\newcommand{\ee}{\end{equation}}
\newcommand{\bea}{\begin{eqnarray}}
\newcommand{\eea}{\end{eqnarray}}
\newcommand{\br}{{\bf r}}

\title{Andreev reflection in bilayer graphene}

\author{T.~Ludwig}
\affiliation{
Instituut-Lorentz, Universiteit Leiden, P.O.~Box 9506, 2300 RA Leiden, The Netherlands
}

\date{Mar 18, 2007}

\begin{abstract}

We consider the differential conductance of a normal-superconductor junction in clean bilayer graphene in the framework of the Dirac-Bogoliubov-de\ Gennes equation. A remarkable suppression of the differential conductance at voltages just below the superconducting gap is found. This can be understood in terms of the spinor structures of the electron and hole excitations, in particular, the reflected {\it valence-band} hole being orthogonal to the incoming electron at normal incidence.

\end{abstract}


\maketitle

\section{Introduction}
\label{intro}

Graphene, with its Dirac-like low-lying excitations, displays a number of unusual transport properties\cite{Novoselov_et_al_05,Novoselov_et_al_06,Zhang_et_al_05}. These result from the pseudospin $1/2$ of the low-lying modes, their linear dispersion, and the vanishing density of states at the Dirac point. The peculiarity that graphene retains a finite conductivity down to the Dirac point\cite{Katsnelson_06,Tworzydlo_et_al_06,Ziegler_06,Nomura_MacDonald_06,Aleiner_Efetov_06,Altland_06,Ostrovsky_Gornyi_Mirlin_06,Titov_06} allows us to explore unusual regimes. Recently normal-superconductor interfaces in graphene have been studied\cite{Beenakker_06,Titov_Beenakker_06,Bhattacharjee_Sengupta_06,Titov_Ossipov_Beenakker_06,Akhmerov_Beenakker_06,Greenbaum_et_al_06,Akhmerov_Beenakker_06b}, where the possibility of the Fermi energy being smaller than the superconducting gap leads to a number of new phenomena based on specular\cite{Beenakker_06} Andreev reflection\cite{Andreev_64}.

In contrast to the monolayer, the low-lying excitations of bilayer graphene are gapless {\it massive} chiral fermions with pseudospin $1$.\cite{McCann_Falko_06,Snyman_Beenakker_06,McCann_06,Ezawa_06} This does not lead to such dramatic effects as the absence of backscattering in the monolayer,\cite{Ando_Nakanishi_98,Ando_Nakanishi_Saito_98,Suzuura_Ando_02,Cheianov_Falko_06} which results, e.g., in Klein tunneling \cite{Katsnelson_Novoselov_Geim_06} or perfect electron-hole conversion at a normal-superconductor (NS) interface at normal incidence and all energies below the superconducting gap\cite{Beenakker_06}. On the contrary, the pseudospin $1$ of the bilayer modes is often difficult to connect with observable quantities.

The purpose of this paper is to examine Andreev reflection at an NS interface in the bilayer and to point out observable effects of the pseudospin $1$ of the low-energy excitations in a way to complement the recently found\cite{Beenakker_06} results for the monolayer. As will be pointed out below, one key feature is the interplay of the valence-band hole mode, which for a given direction has a (two-component) spinor structure {\it orthogonal} to the electron mode, with the other contributing modes.

\section{Model}
\label{sec2}

\begin{figure}[t]
\includegraphics[width=0.8\linewidth]{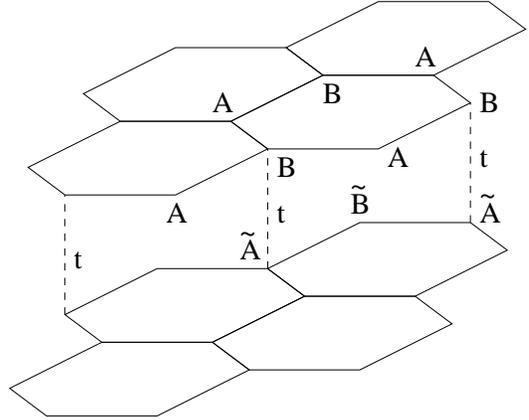}
\caption{\label{layers} The graphene bilayer described by Eq.~(\ref{H}).}
\end{figure}

We consider a clean infinitely wide interface between a normal (N, filling the half plane \mbox{$x>0$}) and superconducting (S, filling \mbox{$x<0$}) region of bilayer graphene.
The bilayer consists of two graphene layers with inequivalent sites $A$ and $B$ in one layer and $\tilde{A}$ and $\tilde{B}$ in the other layer, with $\tilde{A}B$ stacking (see Fig.~\ref{layers}).
In the following we describe it by the Hamiltonian\cite{Wallace_47,McCann_Falko_06,Nilsson_et_al_06}
\be
H=\hbar v
\left(
\begin{array}{cccc}
U     & k-iq  & 0      & 0   \\
k+iq  & U     & t      & 0   \\ 
0     & t     & U      & k-iq\\
0     & 0     & k+iq   & U   \\
\end{array}
\right)
\label{H}
\ee
with the momentum components \mbox{$\vec{k}=\left(k,q\right)$}, where $k$ is normal to the interface and $q$ is parallel to the interface, the hopping element $t$ between (nearest-neighbor) $\tilde{A}$ and $B$ sites, the chemical potential $U$, and the Fermi energy of the normal region $E_F$. The factor \mbox{$\hbar v$}, with $v$ the velocity of the monolayer modes, will be set to $1$ in the following.

As will be discussed below, the four-component Hamiltonian (\ref{H}) is needed to correctly describe the coupling to the strongly doped region\cite{Snyman_Beenakker_06}. This is in contrast to most other situations, where the second-order two-component Hamiltonian
\be
H=\frac{1}{2m}
\left(
\begin{array}{cc}
U                     & -\left(k-iq\right)^2 \\
-\left(k+iq\right)^2  & U     
\end{array}
\right)\:\:,
\label{H2}
\ee
taking into account only the wavefunction amplitudes on sites $A$ and $\tilde{B}$, can be used.\cite{Novoselov_et_al_06,McCann_Falko_06,McCann_06,Katsnelson_Novoselov_Geim_06,Ezawa_06,Snyman_Beenakker_06} The advantage of the two-component Hamiltonian is the more direct insight into the pseudospin and the Berry phase $2\pi$ of the low-energy excitations. This picture may (and will) also be employed for more insight in some particular situations in the following calculations, in which the $B$ and $\tilde{A}$ amplitudes can be disregarded.

It is assumed that superconductivity is induced in the region S by covering it with a superconducting electrode,\cite{Beenakker_06} or that it is caused by some other intrinsic mechanism,\cite{Ferrier_et_al_06,Uchoa_CastroNeto_06} so that issues related to the matching of the hexagonal lattice in N to a different lattice in S (see Ref.~\onlinecite{Blanter_Martin_06}) do not arise.
Further, Eq.~(\ref{H}) assumes that there is no layer asymmetry\cite{McCann_06} and also does not consider corrections to the dispersion relation which become important at very small or large energies.\cite{McCann_Falko_06}

In the absence of a magnetic field, we have time-reversal symmetry, and
the electron and hole excitations of the system are described by the
Dirac-Bogoliubov-de Gennes equation\cite{Beenakker_06}
\be
\left(
\begin{array}{cc}
H-E_F & \Delta \\
\Delta^* & E_F-H
\end{array}
\right)
\left(
\begin{array}{c}
u\\
v
\end{array}
\right)
=
\epsilon
\left(
\begin{array}{c}
u\\
v
\end{array}
\right)
\:,
\label{DBdG}
\ee
where $u$ and $v$ are the electron and hole wave functions, each having four components denoting the amplitudes on the four different atoms in a bilayer unit cell,
\be
u=\left(
\begin{array}{c}
u_1(A)\\
u_2(B)\\
u_3(\tilde{A})\\
u_4(\tilde{B})
\end{array}
\right)
\quad
,
\qquad
v=\left(
\begin{array}{c}
v_1(A)\\
v_2(B)\\
v_3(\tilde{A})\\
v_4(\tilde{B})
\end{array}
\right)
\:,
\ee
and \mbox{$\epsilon>0$} the excitation energy above the Fermi energy \mbox{$E_F>0$}.

We assume the normal region to be weakly doped, and the superconducting region to be strongly doped, described by a potential step of height $U_0$,
\be
U=\left\{
\begin{array}{cc}
-U_0\:, & x<0\\
0\:,    & x>0\:\:.
\end{array}
\right.
\ee
For large positive $U_0$, the Fermi wavelength in S is much smaller than in N, and the pair potential reaches its bulk value at a distance from the interface which is small compared to the Fermi wavelength in N. This justifies the step-function model
\be
\Delta=\left\{
\begin{array}{cc}
\Delta_0\:, & x<0\\
0\:,    & x>0\:,
\end{array}
\right.
\ee
where we may choose \mbox{$\Delta_0>0$} real for convenience.
The junction is assumed to be smooth on the scale of the lattice spacing; therefore, the pair potential does not distinguish between the sublattices or the layers.
In the following we will assume the energy scales to satisfy
\be
U_0 \gg t \gg E_F\,,\:\epsilon\,,\:\Delta_0\:\:.
\label{energy_scales}
\ee
(the relations between $E_F$, $\epsilon$, and $\Delta_0$ are not restricted).
Due to the hierarchy (\ref{energy_scales}) it is essential\cite{Snyman_Beenakker_06} to use the first-order four-component Hamiltonian (\ref{H}) instead of the second-order two-component one (\ref{H2}), which is suitable for most problems. To obtain a result which is independent of the details of the modeling of the contacts, we perform the limit \mbox{$U_0\to\infty$}. The Hamiltonian (\ref{H2}), however, corresponds to the limit \mbox{$t\to\infty$}, which may not be taken first. Therefore the part \mbox{$U_0\gg t$} of the hierarchy (\ref{energy_scales}) results in the necessity to use the four-component Hamiltonian (\ref{H}).

\section{Band structure and kinematics}
\label{sec3}

\begin{figure}[t]
\includegraphics[width=1.0\linewidth]{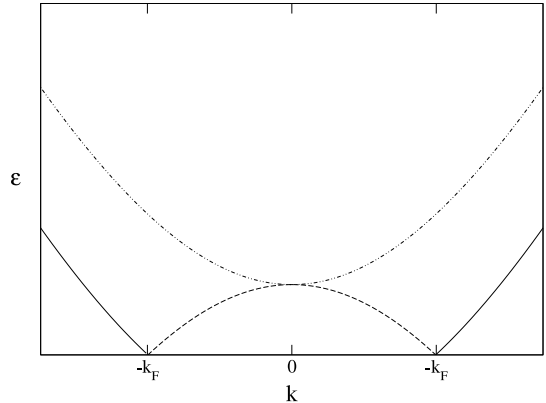}
\caption{\label{bands_N} The band structure of weakly doped (\mbox{$E_F\ll t$}) normal bilayer graphene. At energies below the interlayer hopping $t$, Eq.~(\ref{full_bandstructure}) gives only one electron (solid line) and one propagating hole solution. At excitation energies $\epsilon$ smaller than the Fermi energy $E_F$, the conduction-band hole (dashed line) propagates; at excitation energies above the Fermi energy, the valence-band hole (dot-dashed line) propagates.}
\end{figure}
The band structure of bilayer graphene as described by the Hamiltonian (\ref{H}) is given by
\be
\epsilon+E_F=\pm\frac{t}{2}\pm\frac{1}{2}\sqrt{t^2+4\vec{k}^2}
\label{full_bandstructure}
\ee
and consists of two vertex-touching hyperbolae accompanied by vertically displaced high-energy ones.

In the following we will use the hierarchy of energies (\ref{energy_scales}), which means that in N only the low-energy bands contribute, given the excitation spectrum shown in Fig.~\ref{bands_N},
\be
\epsilon=\left|E_F \pm \left(-\frac{t}{2}+\frac{1}{2}\sqrt{t^2+4\vec{k}^2}\right)\right|\:\:,
\label{low_bandstructure}
\ee
consisting of a propagating electron mode, the conduction-band (CB) hole at \mbox{$\epsilon<E_F$}, and the valence-band (VB) hole at \mbox{$\epsilon>E_F$}. 
The valence band also contributes an evanescent electron mode, which in general is needed to satisfy the continuity conditions at the interface. For the explicit spinor structures of the modes see Appendix~\ref{basis_states}. In the following we approximate the spectrum in N as parabolic, \mbox{$\epsilon,E_F\ll t$}.
The momentum components of the (backscattered) electron are then given by
\be
q=\sqrt{t\left(\epsilon+E_F\right)}\:{\rm sin}\,\alpha\:,
\ee
\be
k=\sqrt{t\left(\epsilon+E_F\right)}\:{\rm cos}\,\alpha\:.
\label{x-momentum}
\ee
At $\epsilon<E_F$ ($\epsilon>E_F$), the CB (VB) hole moves opposite (parallel) to its momentum, while the respective other hole solution is always evanescent.
The CB (VB) hole is outscattered at the angle
\be
\alpha'=
{\rm sign}\left(\epsilon-E_F\right)
{\rm arcsin}\frac{q}{\sqrt{\left|\epsilon-E_F\right|}}\:\:.
\ee
In the limits \mbox{$E_F\gg\Delta_0$} or \mbox{$E_F\ll\Delta_0$}, this leads to retro- and specular Andreev reflection, respectively, just as in the monolayer situation.\cite{Beenakker_06}
If $\alpha>\alpha_c$, with the critical angle $\alpha_c$ given by
\be
\alpha_c=
{\rm arcsin}\,\sqrt{\frac{\left|\epsilon-E_F\right|}{\epsilon+E_F}}\:\:,
\label{critical_angle}
\ee
the hole cannot be matched to the electron at the interface using a real momentum. Instead, $\alpha'$ acquires an imaginary part and both holes become evanescent.

\begin{figure}[t]
\includegraphics[width=1.0\linewidth]{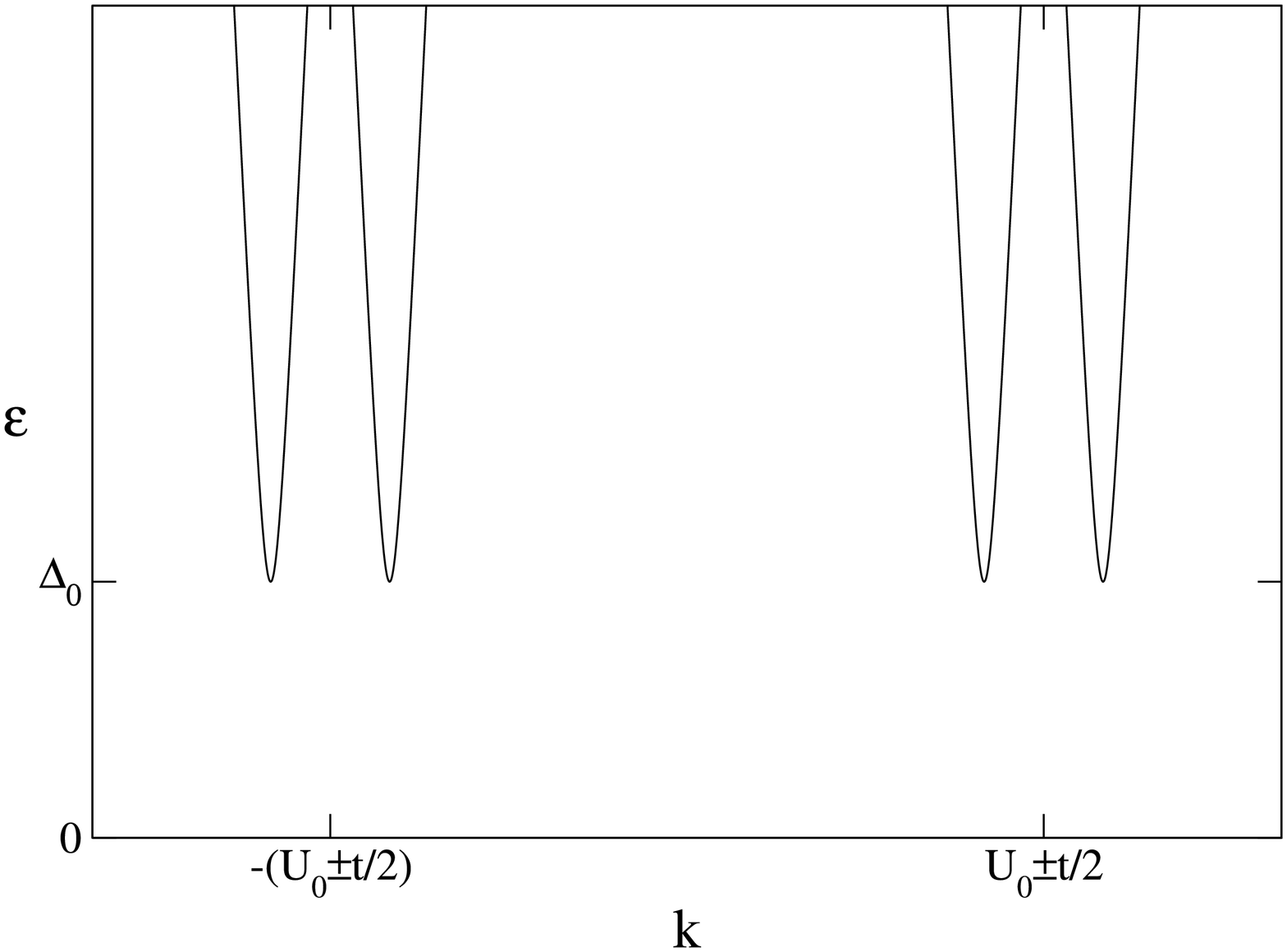}
\caption{\label{bands_S} The band structure of strongly doped (\mbox{$U_0\gg t$}) superconducting bilayer graphene according to Eq.~(\ref{DBdG}) is given by the linear parts of the high- and low-energy dispersion hyperbolae (\ref{full_bandstructure}), with an excitation gap opened up by the pair potential $\Delta_0$.}
\end{figure}

In contrast to the normal region, the superconducting region is assumed to be highly doped, resulting in the excitation spectrum shown in Fig.~\ref{bands_S} which results from the linear parts of all bands. In the limit \mbox{$U_0\to\infty$} the modes in S are normal to the interface.

\section{NS Interface}
\label{sec4}

The reflection coefficients of the NS interface can be found by elementary mode matching demanding continuity of the wave function,
with the boundary conditions that the modes in S decay exponentially or propagate away from the interface.
Equivalently, the effect of the interface may be described by a boundary condition\cite{Titov_Beenakker_06} on the modes in N, which for a strongly doped superconductor acts separately on the two layers (since the modes in strongly doped bilayer graphene are weakly -- symmetrically or antisymmetrically -- coupled monolayer modes),
\be
v(\br)=M\,u(\br)
\label{bc}
\ee
with
\be
M=
\left(
\begin{array}{cc}
{\rm exp}\left\{-i\beta\hat{\bf n}\cdot{\bf\sigma}\right\}&0\\
0&{\rm exp}\left\{-i\beta\hat{\bf n}\cdot{\bf\sigma}\right\}
\end{array}
\right)
\ee
and $\hat{\bf n}$ the unit vector normal to the boundary pointing from N to S.
Eq.~(\ref{bc}) may be used when all excitations in S decay (\mbox{$\epsilon<\Delta_0$}) or move (\mbox{$\epsilon>\Delta_0$}) away from the boundary.
Eq.~(\ref{bc}) is then valid everywhere in S, so by continuity it becomes a boundary condition for the modes in N.

\begin{figure}[t]
\includegraphics[width=0.9\linewidth]{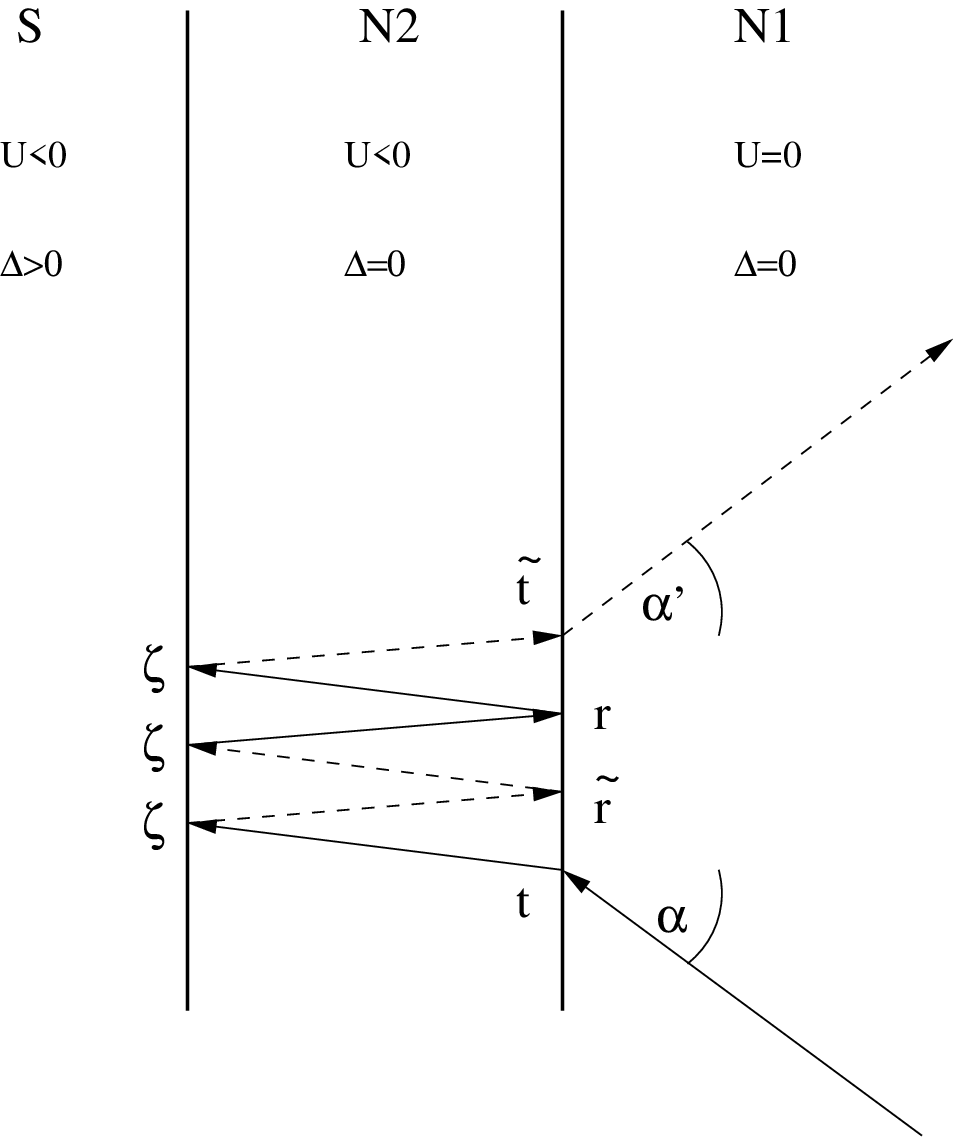}
\caption{\label{NNS} Due to the separation of length scales related to scattering by the chemical potential and by the pair potential, the weakly-doped-N--strongly-doped-S interface may be thought of as being decomposed into an NN interface with the potential step and a strongly doped NS interface without potential step. Scattering processes such as the one shown here are summed up to Eq.~(\ref{two_interfaces}). Solid lines correspond to electron, and dashed lines to hole excitations. The scattering matrices are given in Appendix~\ref{scattering_matrices}.} 
\end{figure}
\begin{figure}[t]
\includegraphics[width=1.0\linewidth]{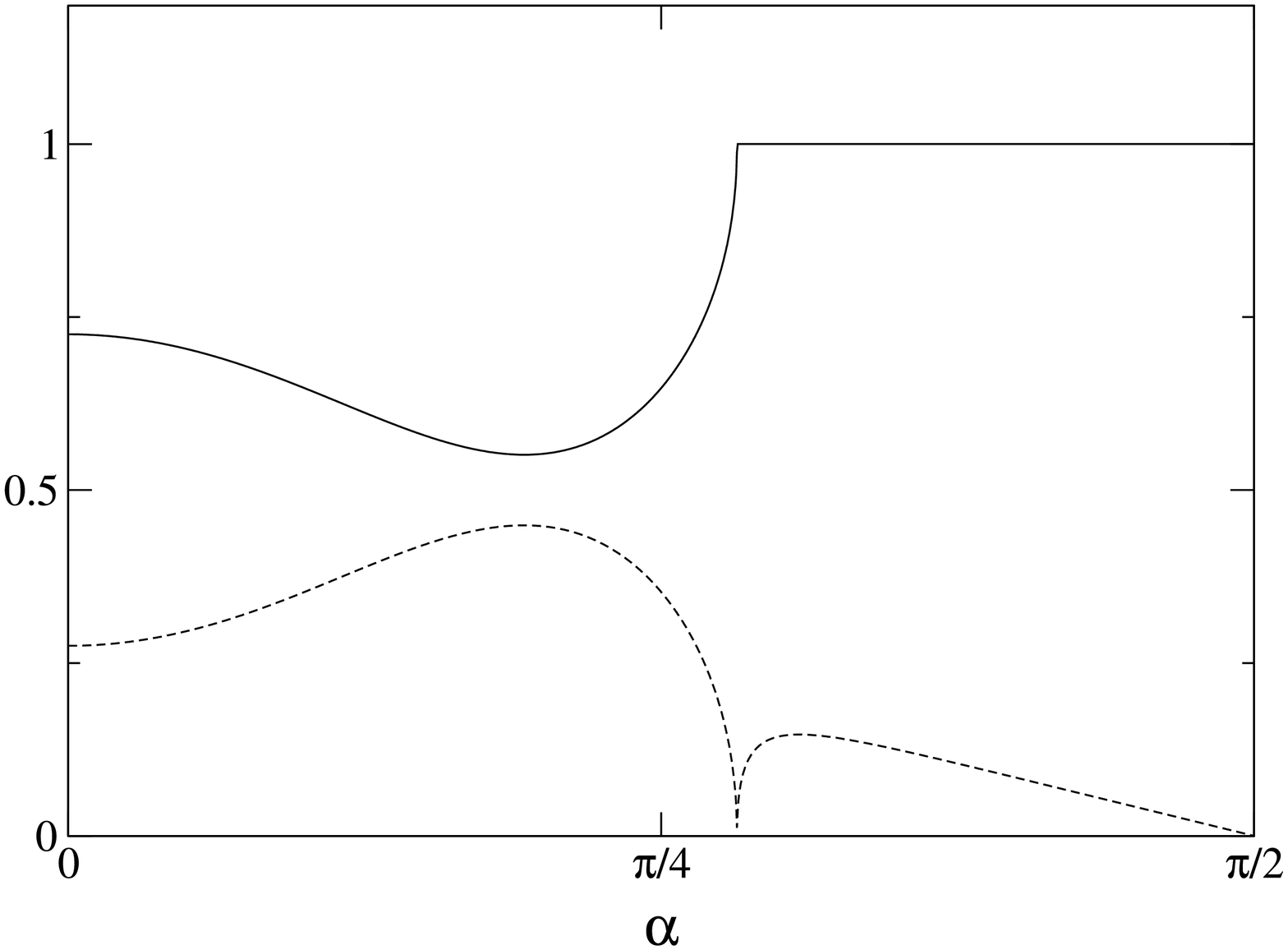}
\includegraphics[width=1.0\linewidth]{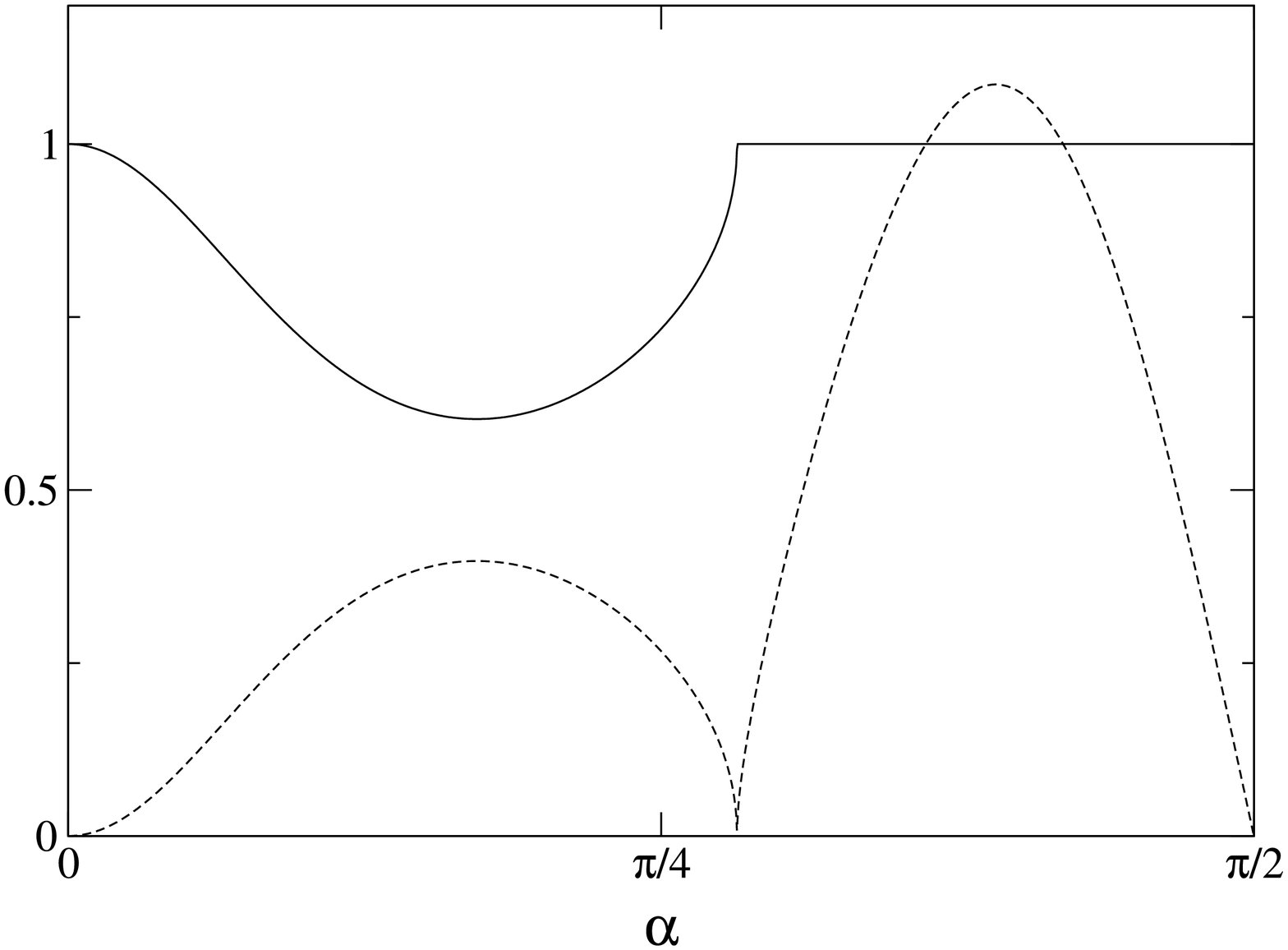}
\caption{\label{above_and_below_EF}
Reflection probabilities for normal and Andreev reflections in the generic situation.
Upper graph: Reflection probabilities $\left|r\right|^2$ (electron, solid line) and $\left|r_A^{\rm CB}\right|^2$ (CB hole, dashed line) {\bf below $E_F$}, plotted against $\alpha$ for $t=10$, $E_F=0.8\,\Delta_0$, and $\epsilon=0.2\,\Delta_0$.
Lower graph:
Reflection probabilities $\left|r\right|^2$ (electron, solid line) and $\left|r_A^{\rm VB}\right|^2$ (VB hole, dashed line) {\bf above $E_F$}, plotted against $\alpha$ for $t=10$, $E_F=0.2\,\Delta_0$, and $\epsilon=0.8\,\Delta_0$.
The probabilities of the propagating modes add up to $1$ (beyond the critical angle $\alpha_c$, the hole is evanescent).
}
\end{figure}

Another way of obtaining the same result arises from the observation that the scattering at the junction can be separated in two effects happening at well separated length scales:\cite{Beenakker_review_97,Akhmerov_Beenakker_06} The scattering by the chemical potential $U$ happens on the scale of the Fermi wavelength of the strongly doped region, while the scattering due to the pair potential $\Delta$ happens on the scale of the superconducting coherence length, which in our model is much longer than the Fermi wavelength.
Therefore, the model described above is equivalent to contacting the weakly doped normal region by a strongly doped normal region, which in turn is in contact to the strongly doped superconductor (see Fig.~\ref{NNS}).
The NN interface is then described by the scattering matrices $S_0(\epsilon)$ for electrons and $\tilde{S}_0(\epsilon)$ for holes, containing reflection and transmission matrices ${\bf r},\tilde{{\bf r}},{\bf t}, and \tilde{{\bf t}}$, respectively. These matrices connect the single propagating mode in the weakly doped region (the parabolic part of the lower one of the dispersion hyperbolae) and the evanescent mode existing in the weakly doped region with the two propagating modes in the strongly doped region (the linear parts of the two hyperbolae).
It is important to note that at \mbox{$\epsilon>E_F$} the relation \mbox{$\tilde{S}_0(\epsilon)=S_0^*(-\epsilon)$} connects the valence band hole scattering matrix with the {\it valence-band} electron scattering matrix, not with the scattering matrix of the incoming conduction-band electron. The 2-spinor structure of the valence band hole is for a given direction orthogonal to the one of the propagating electron. The expressions for $S$ and $\tilde{S}$ are given in Appendix~\ref{scattering_matrices} for the (more interesting) case \mbox{$\epsilon>E_F$}, when the reflected hole is from the valence band.

\begin{figure}[t]
\includegraphics[width=1.0\linewidth]{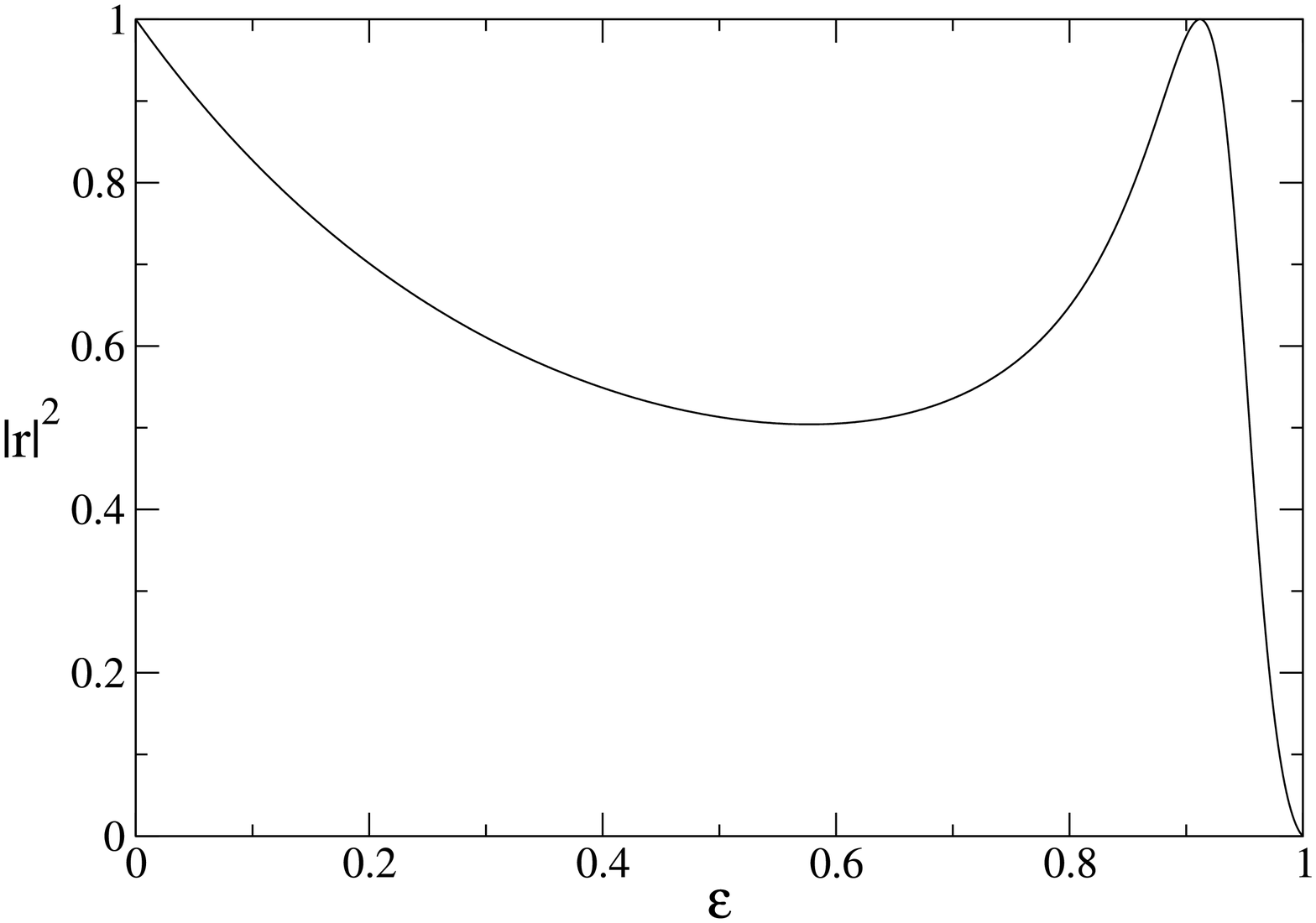}
\includegraphics[width=1.0\linewidth]{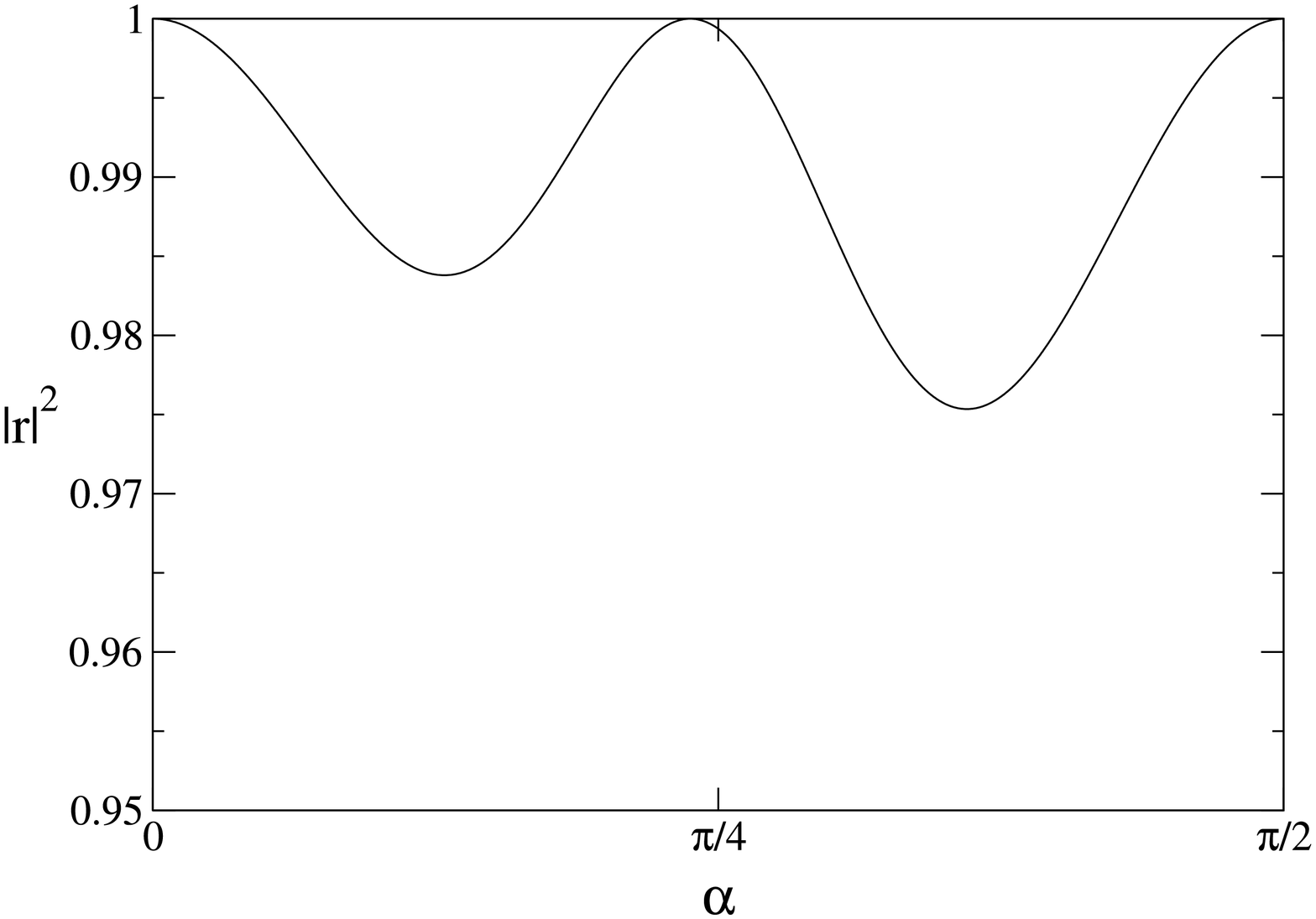}
\caption{\label{dip}
Reflection probability for parameters including a real $\alpha_*$ as given by Eq.~(\ref{special_angle}).
 Upper graph: Plot of the normal reflection probability $|r|^2$ against the excitation energy at the angle of incidence \mbox{$\alpha=\pi/4$} for \mbox{$E_F=0$} and \mbox{$t=3$}. Before falling off to zero (because at \mbox{$eV=\Delta_0$} right-angle reflection is forbidden), $|r|^2$ becomes unity when $\alpha_*$ equals $\pi/4$.
Lower graph:
Plot of $|r|^2$ against the angle of incidence $\alpha$ for \mbox{$E_F=0$}, \mbox{$t=3$}, and \mbox{$\epsilon=0.91\,\Delta_0$} in the special interval given by (\ref{special_energy}). In this interval, $|r|^2$ becomes unity at one more angle $\alpha_*$, given by Eq.~(\ref{special_angle}), other than $0$ and $\pi/2$.
}
\end{figure}

For strong doping, the scattering at the NS interface happens between modes which are normal to the interface. Since \mbox{$U_0\gg t$}, these modes are weakly coupled monolayer excitations. As a result, the subgap scattering matrix of the NS interface is\cite{Beenakker_92,Beenakker_06,Akhmerov_Beenakker_06}
\be
S_A(\epsilon)=
\zeta
\left(
\begin{array}{cc}
0&1\\1&0
\end{array}
\right)\:,\quad
\zeta={\rm exp}
\left\{
-i\,{\rm arccos}\frac{\epsilon}{\Delta_0}
\right\}
\ee
in electron-hole space.
Summing up multiple Andreev reflections between the interfaces in the usual way \cite{Beenakker_review_97} as shown in Fig.~\ref{NNS} results in
\be
r_A(\epsilon,\alpha)=
\zeta\:
\tilde{{ \bf t}}
\left[
{\mathbb I}-\zeta^2\:{\bf r}\:\tilde{{\bf r}}
\right]^{-1}
{\bf t}\:\:.
\label{two_interfaces}
\ee
Note that this is a scalar coefficient since ${\bf r}$ is a \mbox{$2\times 2$} matrix and ${\bf t}$ is a \mbox{$2\times 1$} matrix (see Appendix~\ref{scattering_matrices}).
Carrying out the calculations, the reflection coefficients $r$ for normal reflection and $r_A$, corresponding to Andreev reflection with the propagating hole solution, satisfy \mbox{$\left|r\right|^2+\left|r_A\right|^2=1$} at energies below the gap and are plotted in Figs.~\ref{above_and_below_EF} and \ref{dip}.

\begin{figure}[t]
\includegraphics[width=1.0\linewidth]{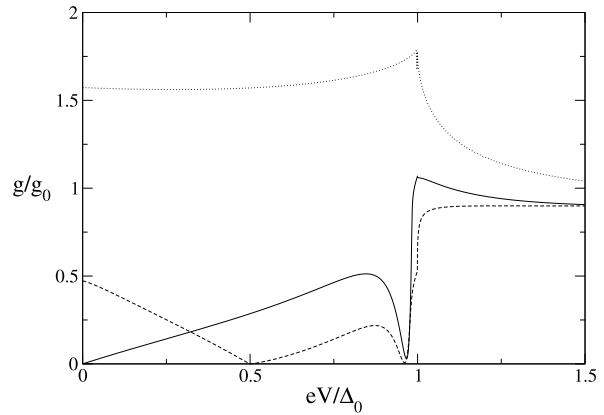}
\caption{\label{plot_g} Plot of the differential conductance of the NS junction, normalized to the ballistic conductance $g_0$, for \mbox{$E_F=0$} and \mbox{$t=10\,\Delta_0$} (solid line), \mbox{$E_F=0.5\,\Delta_0$} (dashed line), and \mbox{$E_F=10\,\Delta_0$} (dotted line). While the dip at \mbox{$eV=E_F$} is straightforward to understand, being due to the absence of available holes, the dip in the range given by Eq.~(\ref{special_energy}) is an effect of the pseudospin $1$ and destructive interference of the two bands, as discussed in the text.}
\end{figure}

The differential conductance of the NS junction can be calculated from the reflection coefficients using the Blonder-Tinkham-Klapwijk formula\cite{Blonder_Tinkham_Klapwijk_82}
\be
\frac{\partial I}{\partial V}=
g_0(V)\int\limits_0^{\pi/2}d\alpha\:{\rm cos}\alpha
\left[
1-\left|r(eV,\alpha)\right|^2+\left|r_A(eV,\alpha)\right|^2
\right]\:\:,
\ee
where $g_0$ is the ballistic differential conductance of a clean sheet of normal bilayer graphene and the reflection coefficient corresponds to the propagating hole solution in N, which, depending on the excitation energy, may be from the conduction band (CB) or from the valence band (VB),
\be
r_A=\left\{
\begin{array}{cc}
r_A^{\rm CB}\:, & \epsilon<E_F\:,\:\alpha<\alpha_c\\
r_A^{\rm VB}\:, & \epsilon>E_F\:,\:\alpha<\alpha_c\\
0\:, & \alpha>\alpha_c\:\:.
\end{array}
\right.
\ee
The results are shown in Figs.~\ref{plot_g} and \ref{plot-different-t}.
The subgap conductance shares with the monolayer result\cite{Beenakker_06} the drop to zero et \mbox{$eV=\Delta_0$}, which just results from the absence of propagating holes at any angle of incidence. In addition, a second dip to almost zero occurs in a voltage range just below the gap. This extra feature is the main result of this paper and is discussed in detail in the following section.

\section{discussion}
\label{sec5}

Unlike the monolayer case, in the bilayer subgap Andreev reflection is always an effect small in $\epsilon/t$. Among the characteristic features of the reflection coefficients is that, at normal incidence, normal reflection always happens with less that unit probability at energies with \mbox{$0<\epsilon<E_F$} but with unit probability at \mbox{$\epsilon>E_F$}. The physical picture for this is the following: The two-spinor ($A\tilde{B}$) structures of the incoming and the reflected electron, as well as of the reflected CB hole, are perfectly compatible (the pseudospin $1$ does not forbid backscattering). Andreev reflection at \mbox{$\epsilon<E_F$} is therefore a correction of order $\epsilon/t$ to this result, arising from the $\tilde{A}B$ components. The reflected VB hole, on the other hand, has a 2-spinor structure {\it orthogonal} to the incoming electron. As a result, there is perfect normal backscattering at \mbox{$\epsilon>E_F$}. This is an effect similar to the perfect Andreev reflection in the monolayer\cite{Beenakker_06} (there, both CB and VB holes have a spinor structure compatible to the incoming electron, but normal reflection of the electron is forbidden).
\begin{figure}[t]
\includegraphics[width=1.0\linewidth]{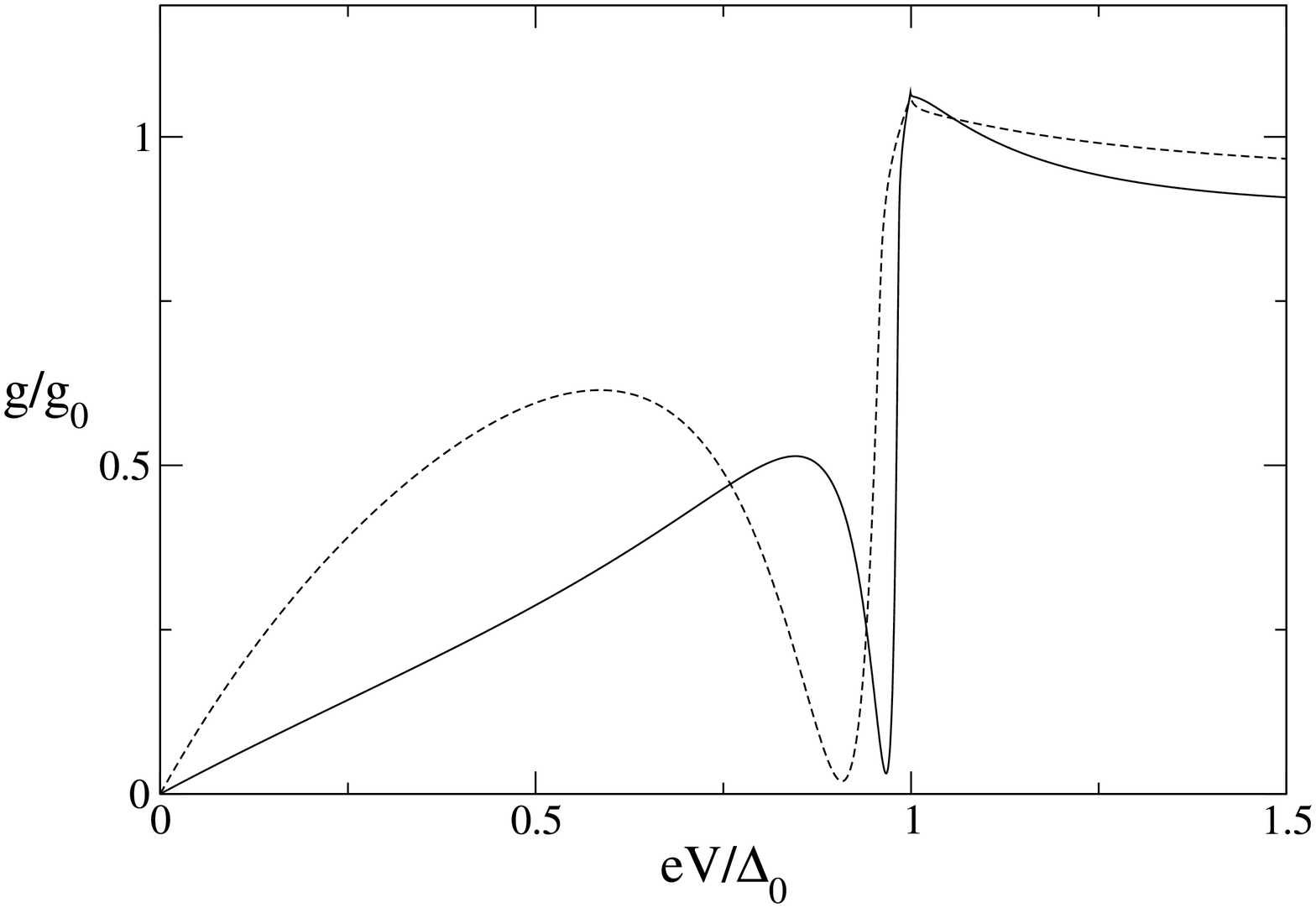}
\caption{\label{plot-different-t} Plot of the differential conductance of the NS junction, normalized to the ballistic conductance $g_0$, for \mbox{$E_F=0$} and \mbox{$t=3\,\Delta_0$} (dashed line) and \mbox{$t=10\,\Delta_0$} (solid line). The characteristic dip in the range given by Eq.~(\ref{special_energy}) moves closer to the gap with increasing $t$. At \mbox{$eV=\Delta_0$}, as seen from Eq.~(\ref{r_at_Delta}), \mbox{$g/g_0=16/15$} (independent of $t$), while at \mbox{$eV<\Delta_0$}, the pointwise limit of the conductance is zero as \mbox{$t\to\infty$}.}
\end{figure}
This effect alone fails to translate directly into a striking feature in the differential conductance, since the faster angular dependence of the spinor structure of the bilayer modes washes out those features upon angular integration.
Under certain conditions, however, perfect normal reflection occurs at yet another angle between $0$ and $\pi/2$.

This happens for \mbox{$E_F=0$} at the angle $\alpha_*$ with
\be
{\rm sin}\alpha_*=\sqrt{\frac{\epsilon/t}{{\rm tan^2\beta}}-1}
\:\:\:,\quad\beta={\rm arccos}(\epsilon/\Delta_0)\:\:,
\label{special_angle}
\ee
for example, at $\alpha_*=\pi/4$ for $t=\sqrt{3}\,\Delta_0$ and $\epsilon=\frac{1}{2}\sqrt{3}\,\Delta_0$ (then $\beta=\pi/6$).
The energy interval in which $\alpha_*$ is real is given by a third-order equation in $\epsilon$; for $E_F=0$ and $\Delta_0\ll t$, it lies close to $\Delta_0$ and is given by
\be
1-\frac{\Delta_0}{2t}\le\frac{\epsilon}{\Delta_0}\le 1-\frac{\Delta_0}{4t}\:\:,
\label{special_energy}
\ee
where \mbox{$\alpha_*=0$} at the lower end of this interval and \mbox{$\alpha_*=\pi/2$} at the upper end.
The complete absence of Andreev reflection at this angle can be understood in the picture of two separate interfaces as the two modes available in the strongly doped normal region interfering destructively.
More insight is obtained by noting that at $\alpha_*$, the boundary condition (\ref{bc}) forms a symmetric dimer on the $\tilde{A}B$ sites if only the CB hole is present (\mbox{$r_A^{\rm VB}=0$}). This boundary condition is satisfied by \mbox{$r=-1$} and the coefficient of the evanescent VB electron mode arising from the condition on the $A$ and $\tilde{B}$ sites, resulting in perfect normal reflection at the angle $\alpha_*$.
The energy range (\ref{special_energy}) thus is the range in which the boundary condition (\ref{bc}) can nontrivially combine the $A$ and $\tilde{B}$ with the $\tilde{A}B$ components of the 4-spinor excitations: at smaller energies, the orders of magnitude are too different to allow for a nontrivial result, leading just to small corrections to the \mbox{$\epsilon/t\to 0$} limit.
On the other hand, at energies above (\ref{special_energy}), the mixing angle $\beta$ (which mixes the components of each of the two layers but does not mix between the layers) becomes too small and the reflection probabilities mirror the physics of a pseudospin $1$ particle scattered from a scalar.\footnote{Since at \mbox{$eV=\Delta_0$} the matrix $M$ in (\ref{bc}) is the unit matrix, we may make this argument neglecting the $B$ and $\tilde{A}$ amplitudes.} In particular, as can be seen from Eq.~(\ref{r_at_Delta}), at \mbox{$\epsilon=\Delta_0$} and \mbox{$\alpha=\pi/4$}, Andreev reflection becomes perfect since scattering off a scalar by $\pi/2$ is forbidden. The scale $t$ thus drops out of the reflection coefficients at \mbox{$\epsilon=\Delta_0$}, so that the conductance at \mbox{$eV=\Delta_0$} depends only on $E_F$. At \mbox{$E_F=0$}, \mbox{$g/g_0=16/15$}.

Since in the energy interval (\ref{special_energy}) the probability of normal reflection reaches unity at three different values in the interval \mbox{$\alpha\in\left[0,\pi/2\right]$}, it deviates very little from unity over the entire range of angles, leading {\it even after angular integration} to the characteristic dip of the differential conductance shown in Figs.~\ref{plot_g} and \ref{plot-different-t}.

The relevance of this result is that it can be linked to the pseudospin of the bilayer excitations, especially the forbidden Andreev reflection into the VB hole at normal incidence, which, due to the finer angular structure of the pseudospin $1$ versus pseudospin $1/2$, does not drastically influence the conductance outside the energy range (\ref{special_energy}).

Experimental observations of superconductivity and Andreev reflections in monolayer graphene contacted by superconducting electrodes have now been reported.\cite{Heersche_et_al_06} The results given here may therefore be expected to be tested in the near future. For the model (\ref{H}) to be applicable, however, superconducting gaps are needed which are larger than the scale below which the Fermi surface breaks up into pockets.\cite{McCann_Falko_06} The main results presented here do not rely on any closer vicinity to the Dirac point other than the requirement that the subgap conductance of the junction is considered. Therefore, they can be expected to be qualitatively valid when there is a connected Fermi surface in a part of the subgap energy range, which requires the superconducting gap $\Delta_0$, the $\tilde{A}B$ hopping $t$, and the velocities $v_3$ and $v$ describing the $AB$ and direct $A\tilde{B}$ hopping, respectively,\cite{McCann_Falko_06} to satisfy\footnote{For example, for the gap $\Delta_0=125\,\mu{\rm eV}$ of Ref.~\onlinecite{Heersche_et_al_06}, a connected Fermi surface below the gap occurs according to Ref.~\onlinecite{McCann_Falko_06} if $v_3/v<0.025$.}
\be
\left(\frac{v_3}{v}\right)^2\lesssim\frac{\Delta_0}{t}\:\:.
\ee
Since, compared to the monolayer, the description of the bilayer is somewhat more idealized, it would be interesting to see how the calculated effect survives in experiment.

\section{Summary}
\label{summary}

In this paper the properties of an NS junction in bilayer graphene have been investigated. The interplay of the pseudospin $1$ excitations of the weakly doped normal region with the strongly-doped superconducting region (which may be regarded as two weakly coupled monolayers) produces a nontrivial $IV$ curve.
While sharing with the monolayer \cite{Beenakker_06} the features arising from the situation \mbox{$E_F\lesssim\Delta_0$}, the most remarkable result is that at voltages just below the gap the differential conductance is strongly suppressed (see Figs.~\ref{plot_g} and \ref{plot-different-t}). This can, although indirectly, be traced back to the pseudospin of the low-energy excitations in the bilayer.

\begin{acknowledgments}
Helpful discussions with
Carlo Beenakker,
Edward \mbox{McCann},
Alexander Ossipov,
Izak Snyman,
and Mikhail Titov
are gratefully acknowledged.
This research was supported by the Dutch Science Foundation NWO/FOM.

\end{acknowledgments}

\appendix

\section{basis states}
\label{basis_states}

In this appendix, we give for reference the electron and hole spinors in the normal region. To carry the same current in $x$ direction, these states should be normalized by the square root of the $x$ components of their group velocities which can be obtained from Eq.~(\ref{x-momentum}). The square roots in these expressions follow the convention that the branch cut lies along the negative real axis and \mbox{$\sqrt{-1}=i$}.

The incoming (left-moving) CB electron is given by
\be
\psi_e^{\rm CB,\,in}=
\left(
\begin{array}{c}
{\rm exp}\{i\alpha\}\\
-\sqrt{\epsilon+E_F}\\
\sqrt{\epsilon+E_F}\\
-{\rm exp}\{-i\alpha\}
\end{array}
\right)
\ee
There are four outscattered modes in N, the reflected CB electron
satisfying \mbox{$k^2+q^2=t(\epsilon+E_F)$},
\begin{widetext}
\be
\psi_e^{\rm CB,\, back}=
\left(
\begin{array}{c}
-{\rm exp}\{-i\alpha\}\\
-\sqrt{\epsilon+E_F}\\
\sqrt{\epsilon+E_F}\\
{\rm exp}\{i\alpha\}
\end{array}
\right)
=
\left(
\begin{array}{c}
-\frac{(k-iq)}{\sqrt{k^2+q^2}}\\
-\sqrt{\epsilon+E_F}\\
\sqrt{\epsilon+E_F}\\
\frac{(k+iq)}{\sqrt{k^2+q^2}}
\end{array}
\right)
=
\left(
\begin{array}{c}
\frac{iq-\sqrt{-q^2+(\epsilon+E_F)}}{\sqrt{\epsilon+E_F}}\\
-\sqrt{\epsilon+E_F}\\
\sqrt{\epsilon+E_F}\\
\frac{iq+\sqrt{-q^2+(\epsilon+E_F)}}{\sqrt{\epsilon+E_F}}
\end{array}
\right)\:\:,
\ee
the VB electron
satisfying $k^2+q^2=-t(\epsilon+E_F)$,
which is always evanescent in our scattering problem,
\be
\psi_e^{\rm VB}=
\left(
\begin{array}{c}
\frac{k-iq}{\sqrt{k^2+q^2}}\\
-i\sqrt{\epsilon+E_F}\\
-i\sqrt{\epsilon+E_F}\\
\frac{k+iq}{\sqrt{k^2+q^2}}
\end{array}
\right)
=
\left(
\begin{array}{c}
\frac{-iq+\sqrt{-q^2-(\epsilon+E_F)}}{\sqrt{-(\epsilon+E_F)}}\\
-i\sqrt{\epsilon+E_F}\\
-i\sqrt{\epsilon+E_F}\\
\frac{iq+\sqrt{-q^2-(\epsilon+E_F)}}{\sqrt{-(\epsilon+E_F)}}
\end{array}
\right)\:\:,
\ee
the CB hole satisfying $k^2+q^2=t(E_F-\epsilon)$,
\be
\psi_h^{\rm CB}=
\left(
\begin{array}{c}
-\frac{k-iq}{\sqrt{k^2+q^2}}\\
-\sqrt{E_F-\epsilon}\\
\sqrt{E_F-\epsilon}\\
\frac{k+iq}{\sqrt{k^2+q^2}}
\end{array}
\right)
=
\left(
\begin{array}{c}
\frac{iq\pm\sqrt{-q^2+(E_F-\epsilon)}}{\sqrt{E_F-\epsilon}}\\
-\sqrt{E_F-\epsilon}\\
\sqrt{E_F-\epsilon}\\
\frac{iq\mp\sqrt{-q^2+(E_F-\epsilon)}}{\sqrt{E_F-\epsilon}}
\end{array}
\right)\:\:,
\ee
where the upper (lower) sign is for the propagating (evanescent) CB hole in order to satisfy the boundary condition that the mode propagates to the right or decays away from the interface;
and finally, there is the VB hole
satisfying $k^2+q^2=t(\epsilon-E_F)$,
\be
\psi_h^{\rm VB}=
\left(
\begin{array}{c}
\frac{k-iq}{\sqrt{k^2+q^2}}\\
-\sqrt{\epsilon-E_F}\\
-\sqrt{\epsilon-E_F}\\
\frac{k+iq}{\sqrt{k^2+q^2}}
\end{array}
\right)
=
\left(
\begin{array}{c}
\frac{-iq+\sqrt{-q^2+(\epsilon-E_F)}}{\sqrt{\epsilon-E_F}}\\
-\sqrt{\epsilon-E_F}\\
-\sqrt{\epsilon-E_F}\\
\frac{iq+\sqrt{-q^2+(\epsilon-E_F)}}{\sqrt{\epsilon-E_F}}
\end{array}
\right)
\:\:.
\ee
\end{widetext}

\section{scattering matrices}
\label{scattering_matrices}

In this appendix, we give the explicit expressions for the scattering matrices of the weakly doped to strongly doped N and the strongly doped N to strongly doped S interfaces for energies $\epsilon$ with \mbox{$E_F\le\epsilon\le\Delta_0$}. The propagating hole is thus the valence band hole, which is the time reverse of a {\it valence-band} electron. In the limit of a large potential step \mbox{$U_0\to\infty$}, the modes in the strongly doped region propagate perpendicular to the interface. For ease of notation, $t$ and $E_F$ are not written explicitly.

Performing the calculations by elementary mode matching, we find the CB electron-scattering matrix
\be
S_0(\epsilon)=
\left(
\begin{array}{cc}
r_{11}    & {\bf t}\\
{\bf t}^T & {\bf r}
\end{array}
\right)
=
\left(
\begin{array}{ccc}
r_{11}   & t_{12}^-    & t_{12}^+   \\
t_{21}^- & r_{22}^{--} & r_{22}^{-+}\\
t_{21}^+ & r_{22}^{+-} & r_{22}^{++}
\end{array}
\right)
\ee
[the index $1$ ($2$) denoting the weakly (strongly) doped normal region according to Fig.~\ref{NNS} and the index $+$ ($-$) denoting the upper (lower) band in the strongly doped region]
with components
\begin{widetext}
\be
r_{11}
=
\frac{\epsilon-\sqrt{\epsilon}\,{\rm cos}{\alpha}-i\,{\rm cos}\left[\alpha-i\,{\rm arcsinh\,sin}\alpha \right]+i\sqrt{\epsilon(1-{\rm sin}^2\alpha)}}{-\epsilon-\sqrt{\epsilon}\,{\rm cos}{\alpha}-i\,{\rm cos}\left[\alpha+i\,{\rm arcsinh\,sin}\alpha \right]-i\sqrt{\epsilon(1-{\rm sin}^2\alpha)}}
\ee
\be
t_{21}^-
=
t_{12}^-
=
\frac{\epsilon^{1/4}\sqrt{{\rm cos}\alpha}\,\Bigl(-i-i\,{\rm exp\{2\,arcsinh\,sin}\alpha\}-2\sqrt{\epsilon}\,{\rm exp\{arcsinh\,sin}\alpha\}\Bigr)}{{\rm exp\{arcsinh\,sin}\alpha\}\left(\epsilon+\sqrt{\epsilon}\,{\rm cos}\alpha+i\,{\rm cos}\left[\alpha+i\,{\rm arcsinh\,sin}\alpha\right]+i\sqrt{\epsilon(1+{\rm sin}^2\alpha)}\right)}
\ee
\be
r_{22}^{--}
=
\frac{\epsilon-\sqrt{\epsilon}\,{\rm cos}\alpha-i\,{\rm cos}\left[\alpha+i\,{\rm arcsinh\,sin}\alpha\right]+i\sqrt{\epsilon(1+{\rm sin}^2\alpha)}}{\epsilon+\sqrt{\epsilon}\,{\rm cos}\alpha+i\,{\rm cos}\left[\alpha+i\,{\rm arcsinh\,sin}\alpha\right]+i\sqrt{\epsilon(1+{\rm sin}^2\alpha)}}
\ee
\be
r_{22}^{+-}
=
r_{22}^{-+}=
\frac{-2i\sqrt{\epsilon}\,{\rm sin}\alpha}{\epsilon+\sqrt{\epsilon}\,{\rm cos}\alpha+i\,{\rm cos}\left[\alpha+i\,{\rm arcsinh\,sin}\alpha\right]+i\sqrt{\epsilon(1+{\rm sin}^2\alpha)}}
\ee
\be
t_{12}^+
=
t_{21}^+
=
\frac{2i\,\epsilon^{1/4}\sqrt{{\rm cos}\alpha}\:{\rm sin}\alpha}{\epsilon+{\rm sin}^2\alpha+i\sqrt{\epsilon(1+{\rm sin}^2\alpha)}+{\rm cos}\alpha\left(\sqrt{\epsilon}+i\sqrt{1+{\rm sin}^2\alpha}\right)}
\ee
\be
r_{22}^{++}
=
\frac{\epsilon+\sqrt{\epsilon}\,{\rm cos}\alpha-i\,{\rm cos}\left[\alpha+i{\rm arcsinh\,sin}\alpha\right]-i\sqrt{\epsilon(1+{\rm sin}^2\alpha)}}{\epsilon+\sqrt{\epsilon}\,{\rm cos}\alpha+i\,{\rm cos}\left[\alpha+i{\rm arcsinh\,sin}\alpha\right]+i\sqrt{\epsilon(1+{\rm sin}^2\alpha)}}\:\:.
\ee
\end{widetext}
Direct inspection verifies that $S_0(\epsilon)$ is symmetric and unitary.

As mentioned in the main text,
the VB hole scattering matrix is related to the {\it valence-band} electron-scattering matrix by $\tilde{S}_0(\epsilon)=S_0^*(-\epsilon)$ and thus needs to be calculated separately since the spinor structure of the valence band differs from the spinor structure of the conduction band. We find
\be
\tilde{S}_0(\epsilon)=
\left(
\begin{array}{ccc}
\tilde{r}_{11}   & \tilde{t}_{12}^-    & \tilde{t}_{12}^+   \\
\tilde{t}_{21}^- & \tilde{r}_{22}^{--} & \tilde{r}_{22}^{-+}\\
\tilde{t}_{21}^+ & \tilde{r}_{22}^{+-} & \tilde{r}_{22}^{++}
\end{array}
\right)
\ee
with components
\begin{widetext}
\be
\tilde{r}_{11}
=
\frac{\epsilon-\sqrt{\epsilon}\,{\rm cos}\alpha-i\,{\rm cos}\left[\alpha-i\,{\rm arcsinh\,sin}\alpha\right]+i\sqrt{\epsilon(1+{\rm sin}^2\alpha)}}{-\epsilon-\sqrt{\epsilon}\,{\rm cos}\alpha-i\,{\rm cos}\left[\alpha+i\,{\rm arcsinh\,sin}\alpha\right]-i\sqrt{\epsilon(1+{\rm sin}^2\alpha)}}
\ee
\be
\tilde{t}_{21}^-
=
\tilde{t}_{12}^-
=
\frac{-2i\epsilon^{1/4}\sqrt{{\rm cos}\alpha}\:{\rm sin}\alpha}{\epsilon+{\rm sin}^2\alpha+i\sqrt{\epsilon(1+{\rm sin}^2\alpha)}+{\rm cos}\alpha\left(\sqrt{\epsilon}+i\sqrt{1+{\rm sin}^2\alpha}\right)}
\ee
\be
\tilde{r}_{22}^{--}
=
\frac{\epsilon+\sqrt{\epsilon}\,{\rm cos}\alpha-i\,{\rm cos}\left[\alpha+i\,{\rm arcsinh\,sin}\alpha\right]-i\sqrt{\epsilon(1+{\rm sin}^2\alpha)}}{\epsilon+\sqrt{\epsilon}\,{\rm cos}\alpha+i\,{\rm cos}\left[\alpha+i\,{\rm arcsinh\,sin}\alpha\right]+i\,\sqrt{\epsilon(1+{\rm sin}^2\alpha)}}
\ee
\be
\tilde{r}_{22}^{+-}
=
\tilde{r}_{22}^{-+}=
\frac{-2i\sqrt{\epsilon}\,{\rm sin}\alpha}{\epsilon+\sqrt{\epsilon}\,{\rm cos}\alpha+i\,{\rm cos}\left[\alpha+i\,{\rm arcsinh\,sin}\alpha\right]+i\sqrt{\epsilon(1+{\rm sin}^2\alpha)}}
\ee
\be
\tilde{t}_{12}^+
=
\tilde{t}_{21}^+=
\frac{\epsilon^{1/4}\sqrt{{\rm cos}\alpha}\Bigl(i+i\,{\rm exp}\{2\,{\rm arcsinh\,sin}\alpha\}+2\sqrt{\epsilon}\,{\rm exp\{arcsinh\,sin}\alpha\}\Bigr)}{{\rm exp\{arcsinh\,sin}\alpha\}\left(\epsilon+\sqrt{\epsilon}\,{\rm cos}\alpha+i\,{\rm cos}\left[\alpha+i\,{\rm arcsinh\,sin}\alpha\right]+i\sqrt{\epsilon(1+{\rm sin}^2\alpha)}\right)}
\ee
\be
\tilde{r}_{22}^{++}
=
\frac{\epsilon-\sqrt{\epsilon}\,{\rm cos}\alpha-i\,{\rm cos}\left[\alpha+i\,{\rm arcsinh\,sin}\alpha\right]+i\sqrt{\epsilon(1+{\rm sin}^2\alpha)}}{\epsilon+\sqrt{\epsilon}\,{\rm cos}\alpha+i\,{\rm cos}\left[\alpha+i\,{\rm arcsinh\,sin}\alpha\right]+i\sqrt{\epsilon(1+{\rm sin}^2\alpha)}}\:\:.
\ee
\end{widetext}
As $S_0$, also $\tilde{S}_0(\epsilon)$ is symmetric and unitary.
Inserting the above expressions into Eq.~(\ref{two_interfaces}) gives the results plotted in the figures of the main text.

While the corresponding analytical expressions for $r$ and $r_A$ in the general case are too lengthy to be displayed here, the following limiting cases can be given:

(1) For $E_F=0$, leading order in \mbox{$\epsilon/t\ll 1$} (the scale $\Delta_0$ drops out):
\bea
r&=&1-2\,{\rm sin}^4\alpha+2i\,{\rm sin}^2\alpha\sqrt{1-{\rm sin}^4\alpha}\:\:,\quad |r|^2=1\:,\nonumber\\
& &\\
r_A^{\rm CB}&=&0\:,\\
r_A^{\rm VB}&=&0\:.
\eea

(2) For $E_F=0,\:\epsilon=\Delta_0$ [this means $\beta=0$ and $M=\mathbb{I}$, now the scale $t$ drops out since $M$ does not mix the large ($A\tilde{B}$) and small ($\tilde{A}B$) components of the 4-spinors]:
\bea
r&=&\frac{-i\,{\rm cos}2\alpha}{-i\,{\rm sin}^2\alpha+\sqrt{1-{\rm sin}^4\alpha}}\:,\:|r|^2={\rm cos}^22\alpha\:,\qquad
\label{r_at_Delta}\\
r_A^{\rm CB}&=&\frac{-1+{\rm sin}^2\alpha-i\sqrt{1-{\rm sin}^4\alpha}}{-i\,{\rm sin}^2\alpha+\sqrt{1-{\rm sin}^4\alpha}}\:,\\
r_A^{\rm VB}&=&\frac{2i\,{\rm cos}\alpha\,{\rm sin}\alpha}{{\rm sin}^2\alpha+i\sqrt{1-{\rm sin}^4\alpha}}\:.
\eea

\end{document}